\documentclass{article}%
\usepackage{graphicx}
\usepackage{amsmath}%
\usepackage{amsfonts}%
\usepackage{amssymb}

\begin{document}

\title{Space-time and Probability}
\author{S. W. Saunders\\Linacre College, University of Oxford}
\maketitle

\begin{center}
\bigskip{\Huge Space-Time and Probability}

{\huge Simon Saunders\footnote{{\huge Philosophy Centre, 10 Merton St., Oxford
OX1 4JJ. simon.saunders@philosophy.ox.ac.uk.
http://users.ox.ac.uk/\symbol{126}lina0174/Saunders.html}}}
\end{center}

\noindent Special relativity is most naturally formulated as a theory of
spacetime geometry, but within the spacetime framework probability appears to
be a purely epistemic notion. It is possible that progress can be made with
rather different approaches - covariant stochastic equations, in particular -
but the results to date are not encouraging. However, it seems a non-epistemic
notion of probability can be made out in Minkowski space on Everett's terms.

I shall work throughout with the consistent histories formalism. I\ shall
start with a conservative interpretation, and then go on to Everett's.

\section{Probability in Consistent Histories}

In the consistent histories approach histories are represented by ordered
products of Heisenberg-picture projection operators, of the form%

\begin{equation}
C(\underline{\alpha}(n,-m))=P_{\alpha_{n}}(t_{n})...P_{\alpha_{1}}%
(t_{1})P_{\alpha_{0}}(t_{0})P_{\alpha_{-1}}(t_{-1})...P_{\alpha_{-m}}(t_{-m}).
\end{equation}

\noindent Here \underline{$\alpha$}($n,-m)$ is the ordered sequence of
variables $<\alpha_{n},...\alpha_{-m}>$; its values are \textit{histories}.
(As is usual, variables will on occasion stand for values as well. I shall
also write \underline{$\alpha$}, \underline{$\alpha$}$^{\prime}$, \ etc. where
beginning and end points of histories do not need to be made explicit.) Each
variable $\alpha_{k},$ where $k\in\{n,...-m\},$ ranges over subspaces of
Hilbert space as defined by a resolution of the identity, i.e. a pairwise
disjoint set of projections $\{P_{a_{k}}(t_{k})\}$ which sum to the identity.
Histories, therefore, are ordered sequences of sub-spaces of Hilbert space.
The question of which set of projections is to be selected at each time is a
version of \textit{the preferred basis problem }of quantum mechanics. I shall
come back to this presently; for the time being, for the sake of concreteness,
suppose the spectrum of each projection is a subset of the configuration space
of the system (call it an \textit{outcome). }The probability of the history
$\underline{\alpha}(n,-m)$ given state $\rho$ (which may be pure or mixed) is:%

\begin{equation}
\Pr(\underline{\alpha}(n,-m))=Tr(C(\underline{\alpha}(n,-m))\rho
C(\underline{\alpha}(n,-m))^{\ast}).
\end{equation}

\noindent This quantity can be obtained by repeated application of Luders'
rule, supposing the projections of Eq.(1) to be measured sequentially,
conditionalizing the state on the outcome in each case.

The conditional probability of history $\ \underline{\alpha}(n,1)$, given
history $\underline{\alpha}(0,-m)$ (``future outcomes conditional on the
present and the past'') - assuming the latter has non-zero probability - is:%

\begin{equation}
\Pr(\underline{\alpha}(n,1)/\underline{\alpha}(0,-m))=\frac{Tr(C(\underline
{\alpha}(n,1))C(\underline{\alpha}(0,-m))\rho C(\underline{\alpha
}(0,-m))^{\ast}C(\underline{\alpha}(n,1))^{\ast})}{Tr(C(\underline{\alpha
}(0,-m))\rho C(\underline{\alpha}(0,-m))^{\ast})}%
\end{equation}

\noindent These quantities are real numbers in the interval $[0,1]$. The sum
of the probabilities for all such histories (holding the conditions fixed) is
equal to one. Neither property holds for the retrospective conditionals, of
the form:%

\begin{equation}
\Pr(\underline{\alpha}(-1,-m)/\underline{\alpha}(0))=\frac{Tr(C(\underline
{\alpha}(0))C(\underline{\alpha}(-1,-m))\rho C(\underline{\alpha
}(-1,-m))^{\ast}C(\underline{\alpha}(0))^{\ast})}{Tr(C(\underline{\alpha
}(0))\rho C(\underline{\alpha}(0))^{\ast})}%
\end{equation}

\noindent(``past outcomes conditional on the present''). A condition that
ensures that these too are correctly normalized and sum to one is the so
called \textit{consistency condition }(also called the \textit{weak
decoherence condition}) [Griffiths 1986], [Omnes 1988]:%

\begin{equation}
\underline{\alpha}\neq\underline{\alpha}^{\prime}\Rightarrow Tr(C(\underline
{\alpha})\rho C(\underline{\alpha}^{\prime})^{\ast})+Tr(C(\underline{\alpha
}^{\prime})\rho C(\underline{\alpha})^{\ast})=0.
\end{equation}

\noindent Consistency is a restriction on the preferred basis. As it stands it
is a weak constraint [Kent and Dowker 1996]; it is more stringent if it is to
hold on variation of the state. Certain variables, and associated spectral
decompositions - integrals of densities that obey local conservation laws, for
example - habitually decohere [Halliwell 1998].

The consistency condition is not needed to ensure normalization and additivity
of probability for histories in the predictive case. We can dispense with it
in the retrodictive case as well; we need only replace the denominator in
Eq.(4) by:%

\begin{equation}
\sum_{<\alpha_{-1}....\alpha_{-m}>}Tr(C(\underline{\alpha}(0,-m))\rho
C(\underline{\alpha}(0,-m))^{\ast}).
\end{equation}

\noindent But this strategy is seriously deficient when we consider the
algebraic structure which histories inherit from their definition. For example
- the one we have at the back of our minds - consider outcomes as subsets of
configuration space. As such they inherit the structure of a Boolean algebra.
One outcome can be contained in another by set inclusion, and this generalizes
naturally to histories:
\begin{equation}
\alpha_{n}\subseteq\alpha_{n}^{\prime},...,\alpha_{-m}\subseteq\alpha
_{-m}^{\prime}\Longleftrightarrow\underline{\alpha}(n,-m))\subseteq
\underline{\alpha}^{\prime}(n,-m).
\end{equation}
The Boolean operations of intersection and union extend to histories as well.
If a probability measure on this space of histories is to respect this Boolean
algebra, then:%

\begin{equation}
\Pr(\bigskip\underline{\alpha}\cup\underline{\alpha}^{\prime})=\Pr
(\underline{\alpha})+\Pr(\underline{\alpha}^{\prime})-\Pr(\underline{\alpha
}\cap\underline{\alpha}^{\prime}).
\end{equation}
This condition is not in general satisfied when probabilities are defined by
Eqs.(1), (2); nor does it follow on introducing normalization factors such as
Eq.(6). Using (1) and (2), from (8) one obtains (5); Eq.(8) is in fact a form
of the consistency condition. It ensures that the probability measure over
histories respects the natural set-theoretic relations that follow from coarse-graining.

In physical terms, an outcome consists in values of certain variables (in our
case, configuration space variables) having values in designated intervals of
reals. On partitioning these intervals one obtains new, finer-grained
outcomes, and new, finer-grained histories. The sums of probabilities for
non-intersecting histories should equal the probability of their union; this
is the consistency condition. The analogous condition is automatically
satisfied by histories in the pilot-wave theory, where - because deterministic
- it is equivalent to an additivity requirement for probabilities of disjoint
subsets of configuration space at a single time.

Consider now the interpretation of probability. Although the notation for
histories, including the time $t_{0},$ is suggestive, it does not in fact
imply that ``the present'' is in any way preferred; it only reflects, let us
say, our particular location in a given history, the one which is actual. We
suppose it is a consistent history. We suppose, moreover, that although the
available data is necessarily approximate, telling us that variables take
values in certain intervals, there is all the same a unique, maximally
fine-grained, history of events - the one which actually occurs. We proceed to
assign probabilities for all such fine-grained histories, conditional on the
coarse-grained history fixed by the available data. The conditional
probabilities of Eq.(3) and (4) are special cases of them. But if the actual
history is maximally fine-grained - no matter that we do not know what it is -
then the probability of every possible outcome at every time, conditional on
the actual history, is either zero or one. The only non-trivial notion of
probability available can only concern incompletely specified histories,
histories which can be further fine-grained (whilst still satisfying the
consistency condition). Probability is epistemic.

This framework as it stands cannot be cast in relativistic form, but there is
a near neighbour to it, where the preferred basis concerns subsets of the
spectra of self-adjoint quantities built out of algebras of local fields. The
various measures and the consistency condition can in turn be expressed in
terms of path integrals [Hartle 1989]. There seems to be no fundamental
obstacle to extending this notion of spacetime probability to relativistic
quantum theory. But the probabilities arrived at in this way are all epistemic.

\section{Probability in the Everett Interpretation.}

\noindent There is no non-epistemic notion of probability consistent with
relativity theory because relativity requires the tenseless spacetime
perspective, and that in turn forces an epistemic notion of probability.

There has been plenty of debate in the philosophy literature about this
argument [Putnam 1966, Maxwell 1986, Stein 1991]; here I wish only to show
that there is certainly a loophole in it - if one is prepared to follow the
logic of the treatment of tense, in moving to the spacetime perspective, to
include the treatment of the determinate and indeterminate as well.

Suppose as before that the present is not in any way privileged - that the
question of what is ``now'', like the question of what is ``here'', is simply
a matter of where one happens to be located (in space-time). Now extend this
analysis to ``determinateness''. We are to view these terms as what
philosophers call \textit{indexicals}, terms whose reference depends on the
context of utterance. But in conformity with relativity, suppose that such
contexts are ineluctably \textit{local}; what is determinate is in the first
instance is what is here and now - and, derivative on this, what has
probability one relative to the here and now. If we are to calculate these
probabilities using Eq.(3),(4) (and supposing our history space is in fact a
quasiclassical domain [Gell-Mann and Hartle 1993]), we will obtain pretty much
the same results as using quantum mechanics as standardly interpreted: events
in the past which would have left a record in the present will be determinate;
events of a similar kind, but in the future, will in general be indeterminate;
events remote from the here and now will likewise, in general, be
indeterminate This, typically, is our determinate vicinity. So one would
expect on general physical grounds.

We arrive, given the preferred basis, not at a space of histories, but at a
space of vicinities. They are all of them possibilities; the question is which
of them are realized. But it would be quite impossible to suppose that only
one of them is; that would be akin to solipsism, an egocentric or at best an
anthropocentric view of reality. No more can there be only a single ``here''.
One might suppose that there is only a single ``now''; there are those who
believe that the past and future are not real, or not in the sense that ``the
now'' is real. But it had better be a global, universal now, or else there is
some local \textit{spatia}l vicinity which has this special status as well -
and we are back to the conflict with special relativity (according to which
there \textit{is} no privileged global notion of ``now''). In non-relativistic
physics we have a half-way house, in which moments but not places can be
singled out as real; it is not available in the relativistic case.

We might link up these possible local vicinities, in spacelike and timelike
directions, and thereby arrive at the set of all possible histories. We would
be led back, that is to say, to the framework previously considered, the
consistent histories formalism - and to the epistemic notion of probability
already considered. The only remaining alternative is to suppose that
\textit{all} possible local states of affairs exist. This I take to be
Everett's approach. All that there is is the set of vicinities, with all the
conditional \ probabilities defined between them. In physics we do not only
take up an \textit{atemporal} perspective, situated at no particular moment in
time, we take up an \textit{acontingent} one, independent of any contingent
event too. But it is a \textit{local }perspective. It says nothing about which
history a particular vicinity is part of; indeed, one and the same vicinity
will be part of many distinct histories - depending on the probabilities.
There is no hint of the epistemic notion of probability that we had before.

We should be clear on the difference between this and a \textit{many histories
i}nterpretation. Reifying all the histories of a given history space does not
materially change the situation regarding the interpretation of probability.
Probability remains as before, epistemic. The actual history is the one in
which one happens to be; probability, as before, concerns the measure of
histories consistent with a given coarse-grained description. It makes no
difference if they are fictitious or real.

What is so different about the Everett approach? It is that there is no
univocal criterion for identity over time. There is no fact of the matter as
to what a given actuality will later turn into. Transition probabilities can
be defined as before (Eqs.(3),(4), where the outcomes now concern local
events), but \ there is no further criterion of identity over time. This
criterion is determinate - one has a unique future event - only insofar as the
probability for that event is one (in which case it is already part of the
vicinity, according to our earlier definition).

It is has been objected by many that the notion of \ probability is incoherent
in the Everett interpretation [Loewer 1996], but one can hardly insist that
for the notion of probability to make any sense the future must already be
settled. The approach is moreover much less extravagant - and stays within the
bounds of what can be locally defined - than the many histories approach,
according to which there are vast numbers of \ qualitatively identical local
vicinities, a different one for every distinct history that can accommodate it
(a point well-known in the philosophy literature [Albert and Loewer 1988]).

But here I am concerned with the intelligibility of the approach, not its
plausibility. Does probability make any sense on its terms? In every other
solution to the problem of measurement, including many-histories, the future
is univocal. There is a unique fact of the matter as what will actually be. It
is by no means clear that this statement can be coherently denied, or that any
meaningful notion of probability can be made out in its absence.

It is not a conceptual necessity, however. It is obviously unnecessary in the
case of \ ordinary objects. Whether the Ship of Theseus, parts of which are
steadily replaced over time, remains really the same, or whether it is the
ship built from its parts that is really the original, has been a worry for
philosophers for millennia. No one else has been much concerned by it. If we
have any very pressing \textit{a priori} convictions on this, it seems they
come into play at the level of \textit{personal} identify; that there must be
a unequivocal notion of identity when we come to \textit{human} beings. But
this too has been the subject of a long and inconclusive debate in philosophy.

The problem can be starkly posed [Saunders 1998]. The human brain has
remarkable bilateral symmetry. It can indeed be divided - an operation known
as a commissurotomy - without obvious deleterious effects (although not as far
as the brain stem and the rectilinear formation). But as shown by a series of
classic experiments [Sperry 1982] the two hemispheres can be made to operate
as distinct and non-communicating centres of consciousness. It is a stretch to
take the further step - and to consider the two hemispheres as physically
separated altogether, so that they may function as separate persons - but it
is hard to see why objections on the grounds of medical feasibility should be
germane to the conceptual difficulty. So let us take this further step. In the
process, suppose too that there is rather greater functional symmetry between
the two hemispheres than is in fact typical. It follows that after the process
of fision there are two perfectly normal persons each with equal right - and
claiming that right - to count themselves the same as the person prior to the
division. Prior to division, it seems there can be no fact of the matter as to
which of the two one is going to be.

How are we to understand this? One can hardly expect to have some sort of
shared identity, that the two persons that result will somehow think in
tandem. It is unreasonable to expect death - each of the two that result will
certainly \textit{say} that they have survived; one would be inclined to call
it survival if only a single hemisphere were to remain. So what does one
expect? On any behavioral criterion, the situation is exactly the same as a
probabilistic event with two possible outcomes. We can speak of probability,
even though there is no fact of the matter as to what one will be.

This is a distinctively philosophical thought experiment, but I do not think
it differs fundamentally from physical thought experiments. Unlike many of the
latter, it may even turn out to be practically possible. And I do not see why
the concept of unequivocal identity over time should be retained as
sacrosanct, when so many other basic concepts of space and time have been so
radically revised in physical science.

There remain of course other difficulties with Everett's ideas - the preferred
basis problem among them. This problem \ is particularly severe for a
one-history interpretation of history space, where the ontology - the question
of what exists - depends on the answer to it. On reflection, for the same
reason, it is just as bad in a many-histories interpretation. It is
ameliorated but not removed in the Everett approach, where the preferred basis
does not determine what ultimately exists, but only the constitution of
particular observers - of which vicinities in the universal state are
habitable [Zurek 1994, Saunders 1993]; but this is a much larger question than
the one that I have been concerned with.

\bigskip

\bigskip

{\LARGE References}

\bigskip

\bigskip

\noindent Albert, D. and B. Loewer [1988] `Interpreting the Many-Worlds
Interpretation', \textit{Synthese }\textbf{77}, 195-213.

\noindent Dowker, F., and A. Kent [1996] `On the Consistent Histories Approach
to Quantum Mechanics', \textit{Journal of Statistical Physics} \textbf{82}, 1575-1646

\noindent Everett III, H. [1957] `Relative State Formulation of Quantum
Mechanics',\textit{\ Reviews of Modern Physics }\textbf{29}, 454-62.

\noindent Gell-Mann, M. and J.B. Hartle [1993] `Classical Equations for
Quantum Systems', \textit{Physical Review }\textbf{D47}, 3345-3382

\noindent Griffiths, R. [1984] `Consistent Histories and the Interpretation of
Quantum Mechanics', \textit{Journal of Statistical Physics}, \textbf{36}, 219-72.

\noindent Halliwell, J. [1998] `Decoherent Histories and Hydrodynamical
Equations', \textit{\ Physical Review D}\textbf{58,} 105015

\noindent Hartle, J. [1989] `The Quantum Mechanics of Cosmology', in
\textit{Quantum Cosmology and Baby Universes: Proceedings of the 1989
Jerusalem Winter School for Theoretical Physics}, ed. S. Coleman, J. Hartle,
T. Piran, and S. Weinberg, World Scientific, Singapore, 1991, pp.65-157.

\noindent Loewer, B.: [1996] `Comment on Lockwood',\textit{\ British Journal
for the Philosophy of Science},\textbf{\ 47}, 229-32.

\noindent Maxwell, N. [1985] `Are Probabilism and Special Relativity
Incompatible?', \textit{Philosophy of Science},\textbf{\ 52}, 23-43.

\noindent Omnes, R.: [1988] \textit{Journal of Statistical Physics},
\textbf{53,} 933.

\noindent Pearle, P. [1990] `Towards a Relativistic Theory of Statevector
Reduction', in A. Miller, ed., \textit{Sixty-Two Years of Uncertainty}, Plenum
Press, New York, p.193-214.

\noindent Putnam, H. [1967] `Time and Physical Geometry', \textit{Journal of
Philosophy}, \textbf{64}, 240-47, reprinted in Philosophical Papers, Vol. 1,
Cambridge University Press, Cambridge, 1975, 198-205.

\noindent Saunders, S. [1993], `Decoherence and Evolutionary Adaptation',
\textit{Physics Letters }\textbf{A184}, 1-5.

\noindent Saunders, S. [1998] `Time, Quantum Mechanics, and Probability',
\textit{Synthese}, \textbf{114}, p.405-44, 1998.

\noindent Sperry, R. [1982] `Some Effects of Disconnecting the Cerebral
Hemispheres', \textit{Science},\textbf{\ 217}, 1223-26.

\noindent Stein, H. [1991] `On Relativity Theory and the Openness of the
Future',\textit{\ Philosophy of Science}, \textbf{58}, 147-67.

\noindent Zurek, W. [1994] `Preferred States, Predictability, Classicality,
and the Environment-Induced Decoherence', in \textit{The Physical Origins of
Time Asymmetry}, J.J. Halliwell, J. Perez-Mercader and W.H. Zurek, eds.,
Cambridge University Press, Cambridge.

:
\end{document}